\documentclass[journal=jpcafh,manuscript=article]{achemso}
\usepackage[utf8]{inputenc}
\usepackage{textcomp}
\usepackage{graphicx}
\usepackage[version=4]{mhchem}
\usepackage{siunitx}
\DeclareSIUnit\au{a.u.}
\usepackage{tabularx}
\DeclareSIUnit\angstrom{\text{\AA}}

\usepackage{comment}
\usepackage{csquotes}
\MakeOuterQuote{"}
\usepackage{epstopdf}
\AppendGraphicsExtensions{.tif}

\title{Theoretical Investigation of The X-Ray Stark Effect in Small Molecules}

\author{Avdhoot Datar}
\email{adatar@smu.edu}
\affiliation{Department of Chemistry, Southern Methodist University, Dallas, TX 75275, USA}

\author{Catherine Wright}
\affiliation{Department of Chemistry, Southern Methodist University, Dallas, TX 75275, USA}

\author{Devin A. Matthews}
\affiliation{Department of Chemistry, Southern Methodist University, Dallas, TX 75275, USA}

\abbreviations{EEF, XAS, XPS, CC}

\begin{document}

\begin{tocentry}
\includegraphics[scale=1]{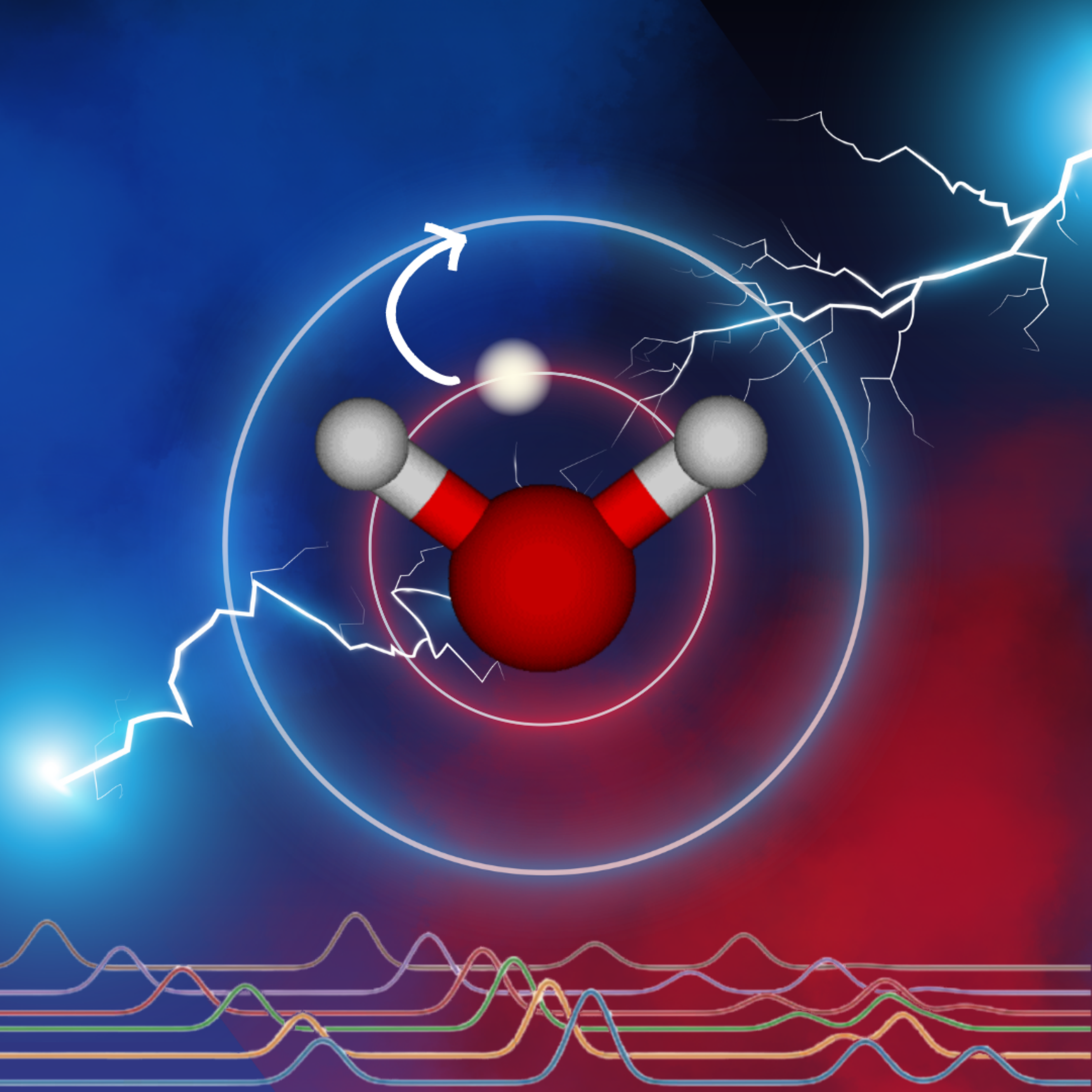} 
\end{tocentry}

\begin{abstract}
We have studied the Stark effect in the soft x-ray region for various small molecules by calculating the field-dependent x-ray absorption spectra. This effect is explained in terms of the response of molecular orbitals (core and valence), the molecular dipole moment, and the molecular geometry to the applied electric field. A number of consistent trends are observed linking the computed shifts in absorption energies and intensities with specific features of the molecular electronic structure. We find that both the virtual molecular orbitals (valence and/or Rydberg) as well as the core orbitals contribute to observed trends in a complementary fashion. This initial study highlights the potential impact of x-ray Stark spectroscopy as a tool to study electronic structure and environmental perturbations at a sub-molecular scale.
\end{abstract}

\section{Introduction}

A central goal in chemistry is achieving selectivity in controlling chemical reactions at the molecular level. A chemical reaction can be controlled by various external factors, such as polarity of solvents, temperature, irradiation with UV--Vis light, external fields, etc. Pioneering computational studies predicted that chemical selectivity can be achieved by applying oriented external electric fields (EEF).\cite{Shaik2004,Hirao2008,DEBIASE2007121} In recent years, there is a growing interest in controlling chemical reactions with EEFs.\cite{Shaik2016,Shaik2020,C8CS00354H,Park2020,Yu2021,Ciampi2018,Che2018,https://doi.org/10.1002/wcms.1438} Numerous experimental works have confirmed the predicted electric field effects on chemical processes.\cite{Zhang2018,Borca2017,Aragones2016} Experimentally, scanning tunneling microscopy (STM) apparatus has been primarily employed for applying the desired electric field.\cite{Zhang2018,Borca2017,Aragones2016}

EEFs alter molecular reactivity by triggering charge separation such that the molecular dipole moment is maximized along the field direction, and hence a reorganization of electron density as well as relaxation of the molecular structure. This phenomenon can lead to a range of effects: alterations of relative stability of tautomers, changes in absorption bands, modifications of various electronic properties (conductor-to-insulator transitions and vice versa, etc.) in the presence of EEFs. Wang et al. showed that evaporation of polythiophene in the presence of an EEF leads to closing of the HOMO-LUMO gap.\cite{Wang2006} Bachler and G{\"{a}}rtner studied the photodissociation of water in the presence of an EEF.\cite{Bachler2016} Dissociation of water and methanol in the presence of EEFs have been studied using density functional theory (DFT) based studies.\cite{PhysRevLett.108.207801,Cassone2014,Cassone2015,Cassone2020} Molecular-scale EEFs can also be generated by environmental effects, for example local EEFs due to charged ions of alkali metal salts lower the water dissociation threshold.\cite{C7CP03663A} Further, it is suggested that sodium iodide (\ce{NaI}) salt would be more efficient than corresponding lithium and potassium analogues in transporting electrons in a dye-sensitized solar cell under biased electric field.\cite{C8CP01155A}

Spectroscopically, an applied EEF gives rise to alterations in the in the energy levels of an atomic or molecular system which is known as the Stark effect. Stark spectroscopy is a general term for describing the spectroscopic changes in the molecular properties due to surrounding electric fields, either laboratory-induced or environmental.\cite{Boxer2009} Generally any molecule interacting with an electrostatic or polarizable environment (solvent, weakly-interacting molecules, etc.) experiences the Stark effect. Thus, intermolecular interactions---hydrogen bonding, van der Waals, $\pi$-stacking, dipole-dipole interactions etc.---can be described in terms of the electric field that is exerted due to such interactions. An advantage of describing these interactions in terms of electric fields is that it provides a unifying terminology for understanding these complex and varying interactions. Towards this end, various efforts are been made to measure the Stark shifts in the electronic transitions\cite{Premvardhan2001,Drobizhev2009,doi:10.1073/pnas.93.16.8362} as well as in vibrational frequencies\cite{Suydam2003,Andrews2000,doi:10.1126/science.1259802}. Computationally, Gurav et al. have studied the electronic stark effect in small molecules using DFT, by calculating UV--Vis spectra of small molecules in the presence of the electric field.\cite{Gurav2018} Donon et al. studied the electronic Stark effect on the $\pi$--$\pi^*$ transition in isolated ion pairs.\cite{Donon2019} Similarly, Stark shifts in vibrational frequencies have been predicted for many molecular systems from computational studies.\cite{Andres1991,MollicaNardo2020,Fried2013,Sowlati-Hashjin2020}

\begin{figure}
\includegraphics[scale=0.25]{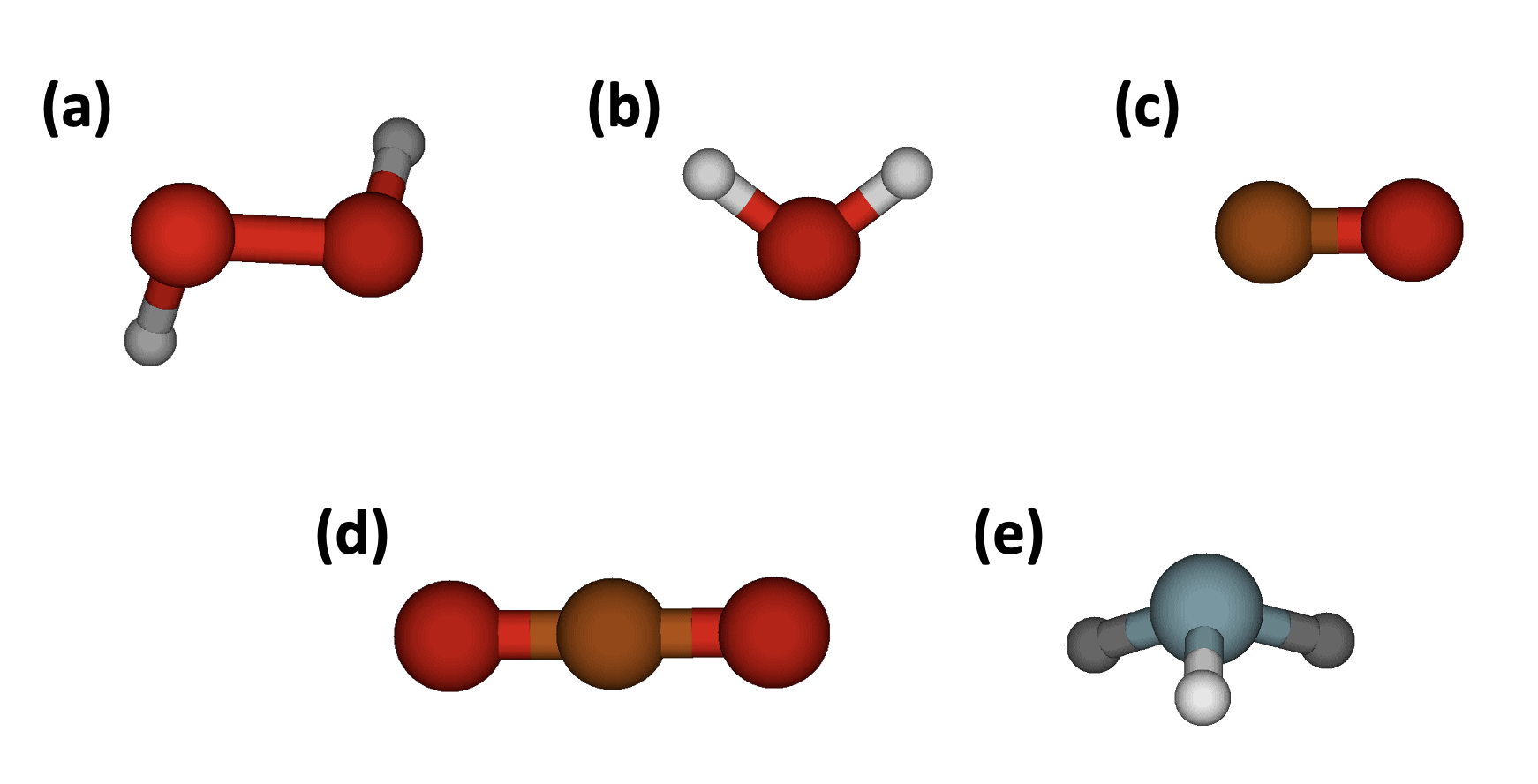}\caption{Prototypical molecules that were chosen to study the x-ray stark effect:
(a) hydrogen peroxide, (b) water, (c) carbon monoxide, (d) carbon dioxide, and (e) ammonia.\label{fig:molecules}}
\end{figure}

Among such spectroscopic techniques, high-resolution x-ray based spectroscopies have emerged as powerful probes for determining the local structure of atoms and molecules, particularly in the soft x-ray regime (roughly 100--\SI{1000}{eV}).\cite{Stohr1992,FADLEY20102,mcneilXrayFreeelectronLasers2010} Since x-ray absorption spectroscopy (XAS) and x-ray photoelectron spectroscopy (XPS) involve excitation of core level electrons, these are excellent techniques for studying the local environment on a sub-molecular scale. The challenges and recent advances in the methods for simulating x-ray spectra are presented in a recent review article by Norman et al.\cite{Norman2018} In this article we are especially interested in the application of equation-of-motion coupled cluster (EOM-CC) techniques, for which many theoretical methods are being to developed to increase the accuracy and efficiency of simulating XAS and XPS of molecules.\cite{corianiCoupledclusterResponseTheory2012,corianiAsymmetricLanczosChainDrivenImplementationElectronic2012,kauczorCommunicationReducedspaceAlgorithm2013,corianiCommunicationXrayAbsorption2015,pengEnergySpecificEquationofMotionCoupledCluster2015,nascimentoSimulationNearEdgeXray2017,zhengPerformanceDeltaCoupledClusterMethods2019,doi:10.1021/acs.jctc.9b00039,doi:10.1080/00268976.2020.1771448,simonsTransitionpotentialCoupledCluster2021,rangaCoreValenceSeparated2021a,Simons2022,e.arias-martinezAccurateCoreExcitation2022} Further, x-ray based spectroscopies provide elemental specificity, which is not possible in vibrational or UV--Vis spectroscopies. Thus, the Stark effect in XAS and XPS could provide a simple technique to study complicated local environmental effects on various molecular species. This article presents our efforts to understand the Stark effect on XAS for a series of small prototypical molecules: hydrogen peroxide (\ce{H2O2}), water (\ce{H2O}), carbon monoxide (\ce{CO}), carbon dioxide (\ce{CO2}), and ammonia (\ce{NH3}), as shown in Figure~\ref{fig:molecules}.

\section{Results}

\begin{figure}
\includegraphics[scale=0.45]{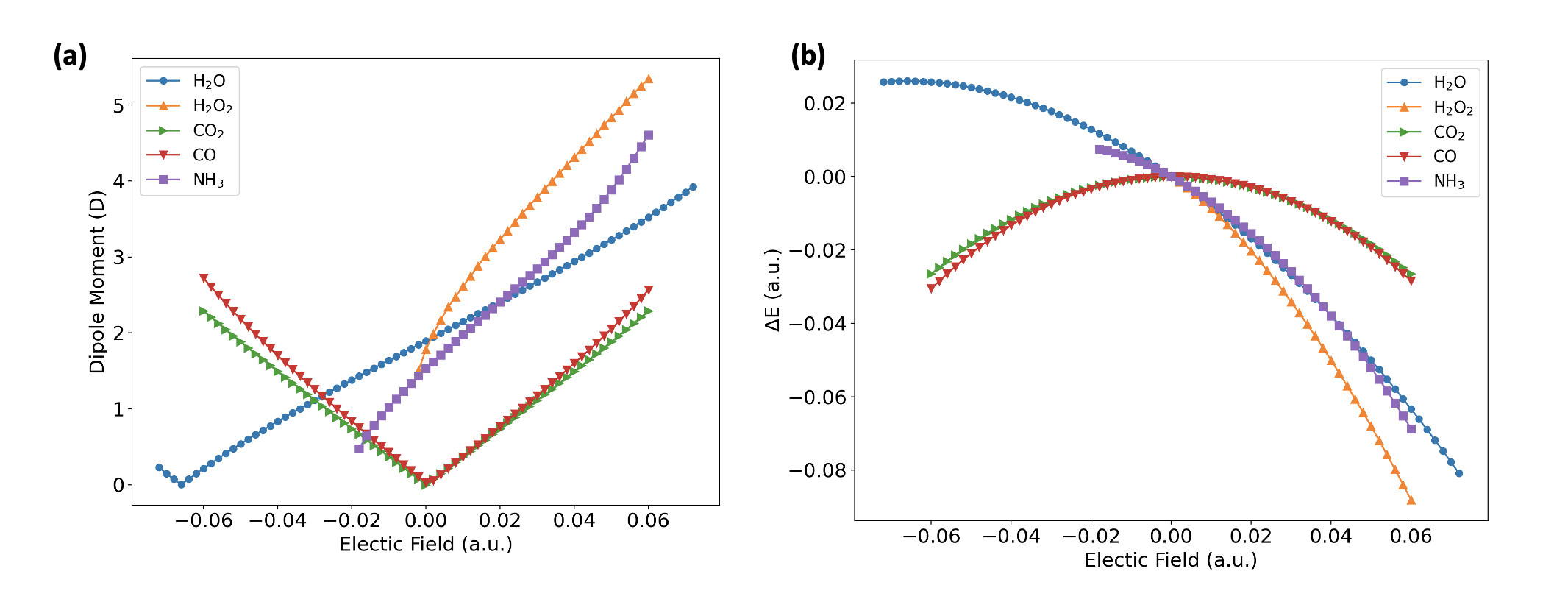}
\caption{Changes in \textbf{(a)} dipole moment, and \textbf{(b)} total ground state electronic energy relative to zero field as a function of the strength of the applied electric field (see text for details).\label{fig:dm-ediff}}
\end{figure}

The Hamiltonian for a molecule with dipole moment ($\hat{\mu}$) interacting with the applied electric field ($\mathcal{E}$) can be written as,
\begin{equation}
\hat{H}=\hat{H}_{0}-\hat{\mu}\cdot\mathcal{E}\label{eq:first}
\end{equation}
Here, $\hat{H}_{0}$ is the field-free Hamiltonian. The direction of electric field applied in this work is parallel or anti-parallel to $\hat{\mu}$. We follow the convention for the direction of the EEF such that a parallel (positive magnitude) electric field results, at first order, in an increase in the molecular dipole moment and a decrease in the total energy with increasing field strength. Relaxation of the molecular geometry and other non-linear effects may result in more complex behavior as is seen in Figure~\ref{fig:dm-ediff}. Uniform electric fields of various field strengths up to \SI{0.06}{\au} (\SI{0.072}{\au} for \ce{H2O}, $\SI{1}{\au} = \SI{51.42}{\V\per\angstrom}$) were applied parallel to to the dipole moments of the isolated molecules optionally followed by relaxation of the molecular geometry in order to minimize the total (field-dependent) energy. For \ce{CO2}, the EEF was applied perpendicular to the molecular axis. At a selection of field strengths and their associated geometries, the evolution of the XAS at the CVS-EOMEE-CCSD level\cite{corianiCommunicationXrayAbsorption2015} was computed.

The observed dipole moment magnitude as a function of electric field strength for relaxed geometries are depicted in Figure~\ref{fig:dm-ediff}a. In each of the examined molecules, the direction of dipole moment is fixed along a conserved symmetry axis; this allows us to easily probe not only fields which a naturally aligned with the dipole moment axis, but "reverse" fields which lead to overall energy raising in most cases. For \ce{CO}, \ce{H2O}, \ce{H2O2}, and \ce{NH3}, a reverse field eventually leads to a non-polar molecular geometry, where the perturbed electronic dipole moment exactly cancels the nuclear dipole moment. The gradient of the molecular energy can be obtained via the Hellman-Feynman theorem\cite{Hellmann1937,PhysRev.56.340} as,
\begin{equation}
dE/d|\mathcal{E}| = -\langle\hat{\mu}\rangle\cdot\mathcal{\hat{E}}\label{eq:ediff}
\end{equation}
where $\mathcal{\hat{E}}$ is the EEF direction.\footnote{Note that \eqref{eq:ediff} is also valid for the coupled cluster energy even through there are distinct left- and right-hand wavefunctions, as long as the dipole moment expectation value is calculated using the "biorthogonal" expression.} This explains the negative (zero) slope for polar (non-polar) molecules at $\mathcal{E}=0$, but also the fact that a maximum in the energy is obtained when the aforementioned dipole moment cancellation occurs. In \ce{H2O} and \ce{NH3} this occurs at rather large field strengths ($\sim\SI{0.065}{\au}$ and $\sim\SI{0.02}{\au}$, respectively), which reflects the relative rigidity of the molecular frame and a moderate polarizability in the field direction. In \ce{H2O2}, the field strength necessary for a zero dipole moment is much lower ($<\SI{0.01}{\au}$), as the \ce{OO} bond rotation is much more facile and and only the modest \ce{OH} bond dipoles must be cancelled out. Additionally, for \ce{NH3} and \ce{H2O2} we encounter field strengths (approximately \SI{0.025}{\au} and \SI{0.006}{\au}, respectively) which are intense enough to cause an inversion of the molecular geometry that essentially "flips" the effective field direction. The fields strengths for inversion and dipole moment cancellation are too close to be distinguished in these latter systems, and we present all points for which we can obtain reliably converged solutions.

While \ce{CO} is polar, is has only a small permanent dipole moment (\SI{0.0275}{D}) which is quickly driven to zero by an opposing EEF. Additionally, the significant polarizability of \ce{CO} causes the molecular dipole moment to change sign (indicating a partial negative charge on C and positive charge on O) for almost all negative field strengths. The magnitude of the induced dipole moment and the resulting energy charge are also rather large given the small change in bond length (see Figure~S1 in the ESI). \ce{CO2} is linear and non-polar, and an EEF applied in any direction perpendicular to the molecular axis, as is done here, results in an identical response. Data for both "positive" and "negative" directions are shown to contrast the \emph{nearly-}symmetric behavior of \ce{CO}. The close similarity of the curves for both molecular dipole moment and total energy change reflect the similar polarizability and weak geometric perturbation (high rigidity) in both systems due to multiple $\pi$ bonds.

With regards to changes in the total ground state energy (Figure~\ref{fig:dm-ediff}b), we observe highly parabolic curves, indicating a field-dependent response dominated by first- and second-order Stark effects. Second-order effects, measured by the curvature of the energy change, seem to be roughly similar in each system, while strong first-order effects are evident in the polar systems. These effects are simply obtained from (2) and by further differentiation, yielding the static dipole moment and anisotropic polarizability as the controlling parameters (see Table S1 and accompanying analysis in the ESI). As previously noted, maxima in the field-dependent energy are obtained when the EEF drives the system into a non-polar configuration. At high field strengths, some systems become moderately ionic and experience significant bond lengthening and angle changes. For example, at a field strength of \SI{0.06}{\au}, the bond angle of \ce{H2O} decreases to \SI{97}{\degree} while the partial charge on O increases from \SI{-0.47}{e} to \SI{-0.8}{e}.

In the following sections, we present field-dependent XAS for each system. The simulated spectra are broadened using a Gaussian profile with a full-width half maximum of \SI{0.2}{eV}. In each case, tentative assignments are given as well as guide lines indicating the evolution of individual bands as a function of field strength. These lines are intended primarily to guide the eye and may differ from a proper analytic continuation of the adiabatic energy levels. The electronic character of individual bands change significantly due to field-dependent mixing, similarly to the non-adiabatic couplings common when examining geometry-dependent potential energy surfaces.\cite{KDC}

\subsection{Hydrogen peroxide}

\begin{figure}
\includegraphics[scale=0.35]{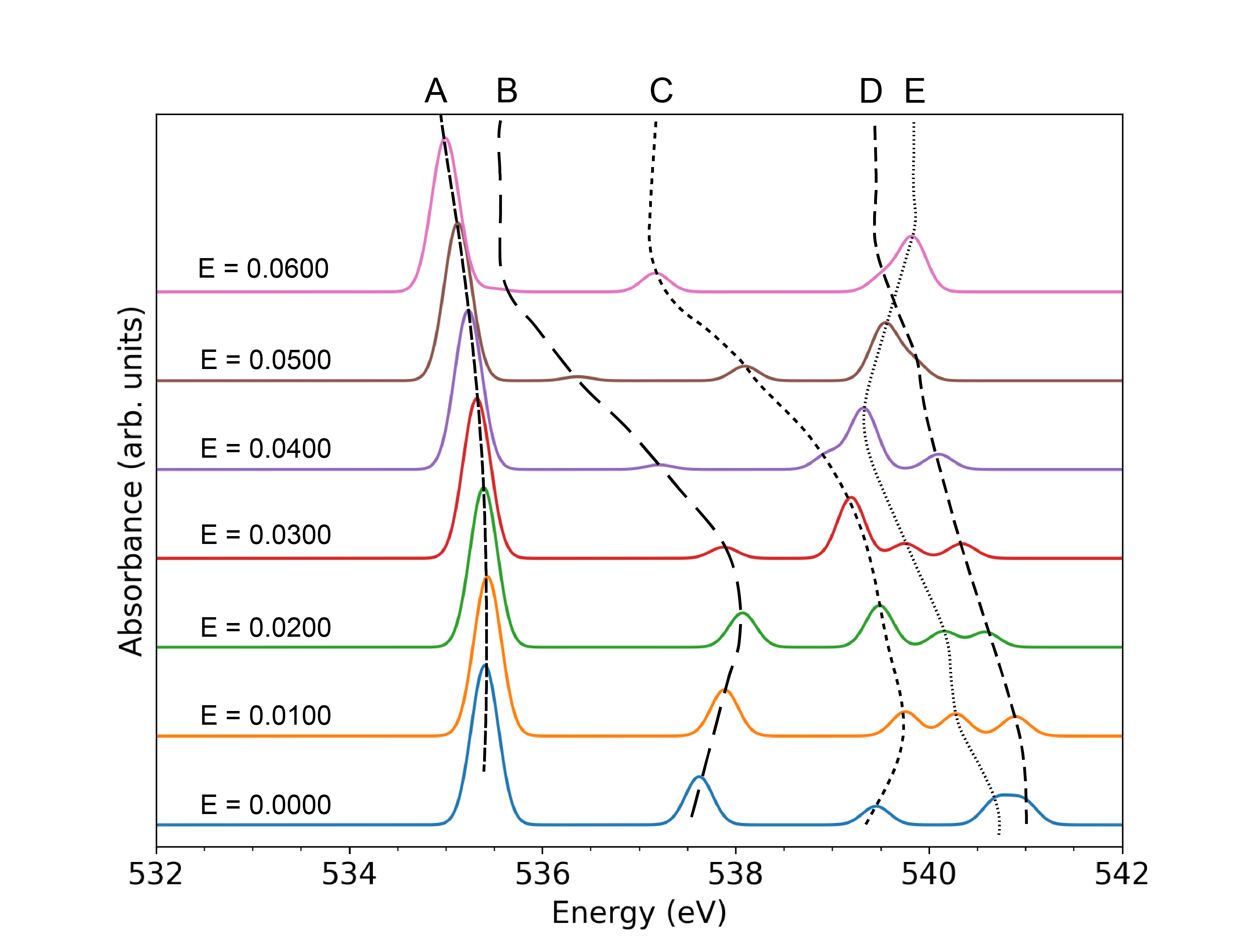}
\caption{\ce{H2O2}: Evolution of the O1s XAS as a function of EEF strength. Corresponding geometries are optimized in the presence of the electric field. Significant absorption bands are indicated by capital letters.\label{fig:XAS-H2O2}}
\end{figure}

\begin{table}
\caption{\ce{H2O2}: NTOs of the O1s core-excited states.\label{tab:H2O2}}

\begin{tabular}{|m{2cm}|m{2cm}|m{4cm}|>{\centering\arraybackslash}m{3cm}|}
\hline 
Band  & Symmetry  & Assignment  & Virtual NTO \\
\hline 
\textbf{A}  & $B$  & $\sigma_{\ce{OO}}^*$  & \includegraphics[scale=0.4]{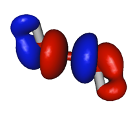}\\
\hline 
\textbf{B}  & $A$  & Rydberg/$\sigma_{\ce{OH}}^*$  & \includegraphics[scale=0.4]{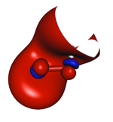}\\
\hline 
\textbf{C}  & $B$ & Rydberg/$\sigma_{\ce{OH}}^*$  & \includegraphics[scale=0.4]{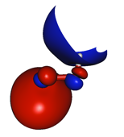}\\
\hline 
\textbf{D}  & $B$  & $3p_{x}$  & \includegraphics[scale=0.4]{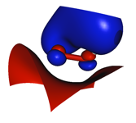}\\
\hline 
\textbf{E}  & $B$  & $3p_{y}$  & \includegraphics[scale=0.4]{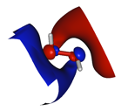}\\
\hline 
\end{tabular}
\end{table}

The XAS for O1s core-excitation of hydrogen peroxide with varying electric field strength is presented in Figure~\ref{fig:XAS-H2O2}, while a summary of significant absorption bands \textbf{A}--\textbf{E}, including a depiction of the virtual transition natural orbitals (NTOs), is presented in Table~\ref{tab:H2O2}. Hydrogen peroxide has two O1s molecular orbitals of $A$ (symmetric) and $B$ (antisymmetric) symmetry. Both transitions are dipole-allowed and contribute to the simulated spectrum as closely-spaced pairs of $A$- and $B$-symmetry transitions. Thus, in Table~\ref{tab:H2O2} we classify the bands by the symmetry of the virtual NTO. The virtual NTO is essentially the same for both components (see ESI).

As the electric field strength is increased from 0 to \SI{0.06}{\au}, the \ce{\angle HOOH} torsion angle is rapidly reduced from its starting value of \SI{113}{\degree} and the molecular geometry approaches a cis-planar structure. Note that a trans-planar structure is instead obtained at a field strength between 0 and \SI{-0.01}{\au}, showing that essentially the full range of torsional geometries is sampled along this profile. Since absorption band \textbf{A} is due to excitation to the $\sigma_{\ce{OO}}^{*}$ orbital, which is almost orthogonal to the applied field, it experiences a minimal shift with changing field strength. On the other hand, the absorption bands \textbf{B} and \textbf{C} experience a significant shift in the absorption energy due to the excitation to $\sigma_{\ce{OH}}^{*}$ orbitals (with significant mixing of Rydberg orbitals). In the fixed-geometry simulation (Figure~S6), band \textbf{A} exhibits essentially no change, showing that observed change in the relaxed case is (almost) entirely geometry-dependent. Actually, with increasing field strength there is a cancellation between two small electronic contributions: a lowering of the core orbital energy (and raising of the core ionization potential), and a concomitant lowering of the $\sigma_{\ce{OO}}^*$ energy due to lengthening of the \ce{OO} bond (Figure~S3). The $\sigma_{\ce{OH}}^{*}$ bands show differing effects due to geometrical relaxation, with \textbf{B} largely unaffected and \textbf{C} blue-shifted by 1--\SI{2}{eV} at high field strengths. This may be due to the combination of the increase in \ce{OH} bond length and the enhanced overlap of the $\sigma_{\ce{OH}}^*$ orbitals as field strength increases: in the symmetric \textbf{A} band, these effects cancel, while in the anti-symmetric \textbf{B} band they both lead to raising of the virtual orbital energy. The Rydberg-dominated bands \textbf{D} and \textbf{E} show moderate change in their absorption energies as a function of field strength, although band \textbf{E} is red-shifted in strong fields relative to the fixed-geometry spectrum, likely due to enhanced mixing with band \textbf{C}. The Rydberg $3p$ orbitals involved in these bands are oriented perpendicular to the field, explaining their weak field dependence. The $3p_z$ Rydberg transition is likely strongly mixed with the valence $\sigma^*$ transitions which do instead depend significantly on the field strength. The features leading to the observed behavior depend mostly on the properties of the valence orbitals rather than the core orbitals. However, the simulated spectra differ from the valence (UV) field-dependent spectra due to the lack of participation of the occupied valence orbitals. Thus, the x-ray Stark spectra for \ce{H2O2} efficiently probe the virtual orbital space only. Modulation of the core orbital energies is limited to 0.1--\SI{0.2}{eV} and is only a minor contribution to the overall shifts.

\subsection{Water}

\begin{figure}
\includegraphics[scale=0.35]{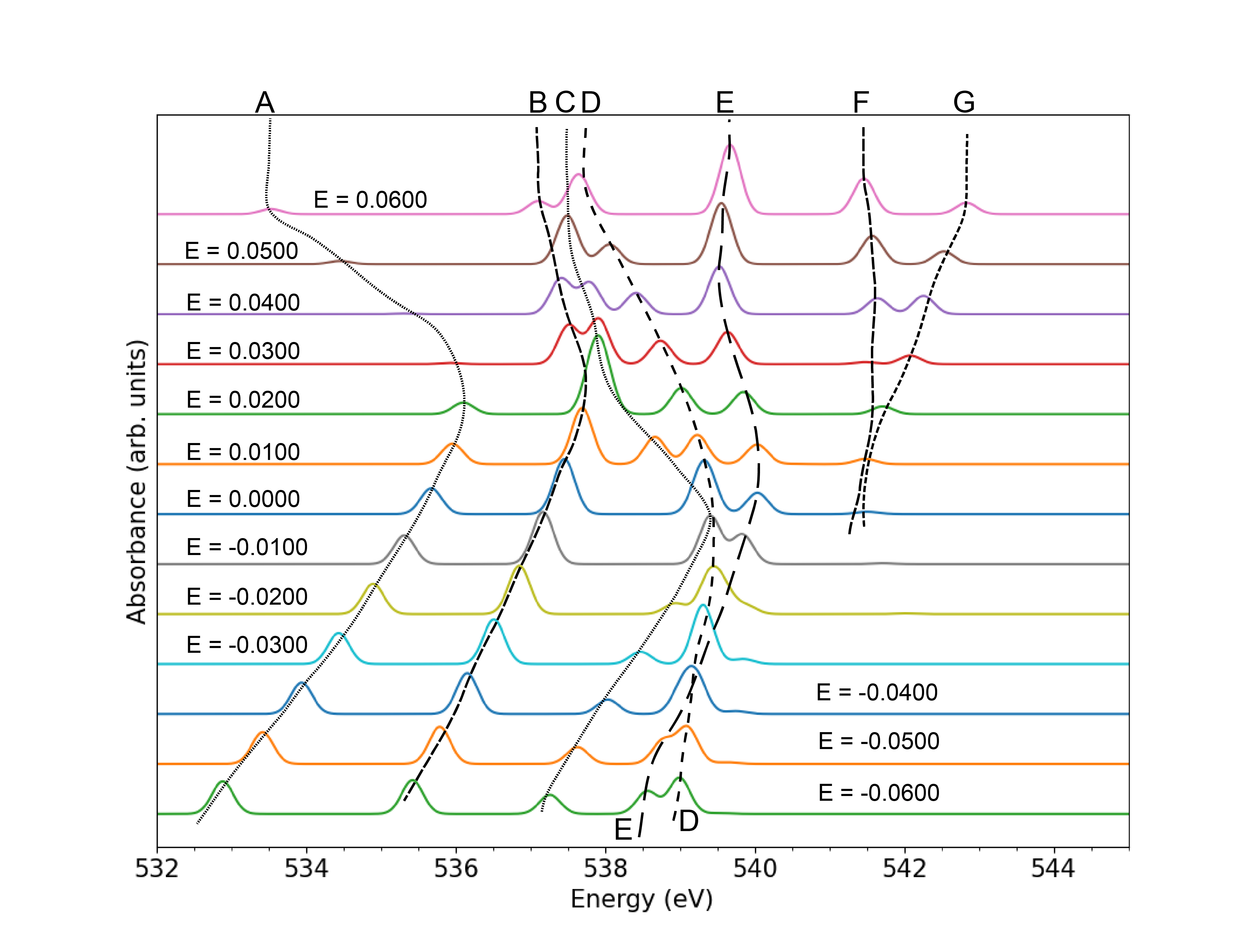}
\caption{\ce{H2O}: Evolution of the O1s XAS as a function of EEF strength. Corresponding geometries are optimized in the presence of the electric field. Significant absorption bands are indicated by capital letters.\label{fig:XAS-H2O}}
\end{figure}

\begin{table}
\caption{\ce{H2O}: NTOs of the O1s core-excited states.\label{tab:H2O}}

\begin{tabular}{|m{2cm}|m{2cm}|m{4cm}|>{\centering\arraybackslash}m{3cm}|}
\hline 
Band  & Symmetry  & Assignment & Virtual NTO\\
\hline 
\textbf{A}  & $A_{1}$  & $3s$  & \includegraphics[scale=0.4]{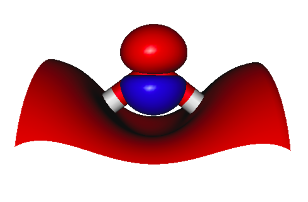}\\
\hline 
\textbf{B}  & $B_{1}$  & $3p_{x}$  & \includegraphics[scale=0.4]{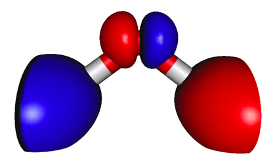}\\
\hline 
\textbf{C}  & $A_{1}$  & $3p_{z}$  & \includegraphics[scale=0.4]{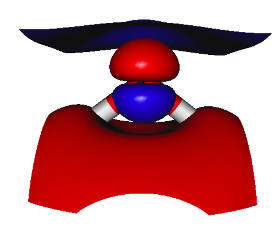}\\
\hline 
\textbf{D}  & $B_{2}$  & $3p_{y}$  & \includegraphics[scale=0.4]{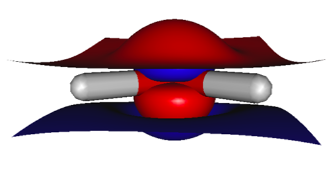}\\
\hline 
\textbf{E}  & $B_{1}$  & $3d_{xz}$  & \includegraphics[scale=0.4]{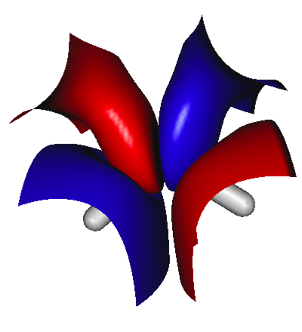}\\
\hline
\end{tabular}
\end{table}

The XAS for the water molecule is studied over a range of electric field strengths from -0.06 to \SI{0.06}{\au}, as depicted in Figure~\ref{fig:XAS-H2O}. The main absorption bands are summarized in Table~\ref{tab:H2O}. The calculated transitions are due to excitation of a core electron to $3s$, $3p$, and $3d$ Rydberg orbitals, without significant participation of the $\sigma^*$ orbitals. The $3p_z$ and $3p_y$ absorption bands are increasingly split as the electric field is increased in the negative direction (opposite to the dipole moment), while the $3p_x$ band is consistently found at lower energy. Interestingly, all $3p$ bands become quite close in energy for strong fields in the positive direction, despite significant geometric anisotropy. On the other hand, inspection of the electronic density isosurfaces (data not shown), highlights a rather spherical electronic distribution about the O atom. The effect of this electronic configuration is also seen in the in intensity of band \textbf{A}: since a spherical electronic density leads to atomic selection rules, this effectively "$S \gets S$" transition becomes dipole-forbidden in the range of field strengths from 0.03 to \SI{0.04}{\au}. 

Band \textbf{E}, which may gain some intensity due to mixing with band \textbf{B} ($3p_x$) of the same symmetry, also closely follows the same energetic trend with field strength as \textbf{B}. At positive field strengths band \textbf{E}, along with bands \textbf{F} and \textbf{G} of indeterminate $3d$ character, gain significant intensity, with band \textbf{E} becoming the most intense at a field strength of \SI{0.06}{\au}. The bands \textbf{D} and \textbf{E} are close in absorption energy for negative electric fields, but are separated in the positive electric field direction. The absorption bands \textbf{E} gains intensity at higher positive electric field strength due to mixing of the $3d_{xz}$ orbital with the $\sigma_{\ce{OH}}^{*}$ orbital of the same symmetry.

From the comparison with XAS simulated at the fixed, field-free optimized geometry (as shown in Figure~S2 of the ESI), it is understood that the effect of change in molecular geometry on the Stark shifts are minimal for the case of \ce{H2O}. In this case the Stark shifts mainly depend on the change in electron density of due to application of the electric field. 

\subsection{Ammonia}

\begin{figure}
\includegraphics[scale=0.35]{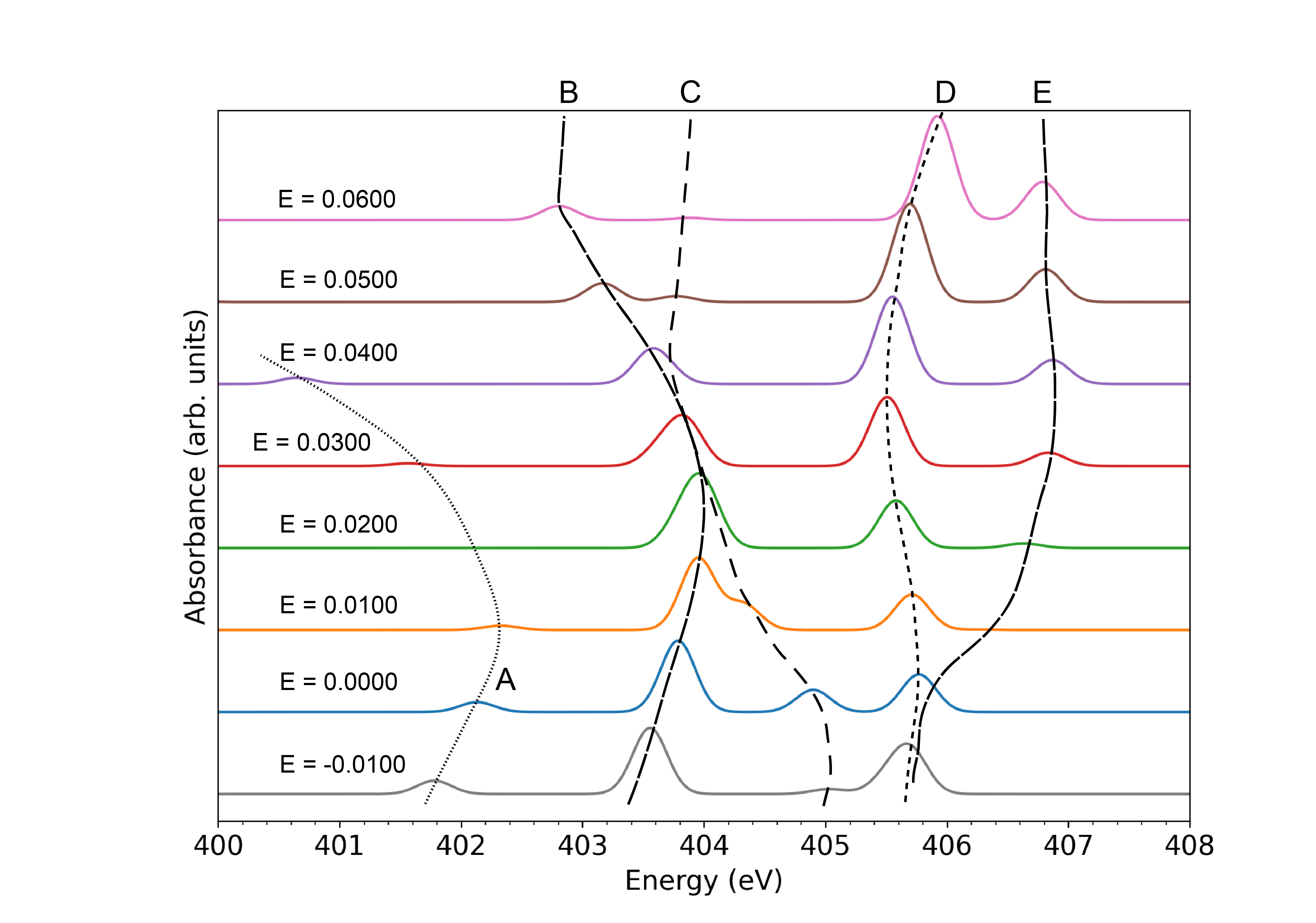}

\caption{\ce{NH3}: Evolution of the N1s XAS as a function of EEF strength. Corresponding geometries are optimized in the presence of the electric field. Significant absorption bands are indicated by capital letters.\label{fig:XAS-NH3}}
\end{figure}

\begin{table}
\caption{\ce{NH3}: NTOs of the N1s core-excited states.\label{tab:NH3}}

\begin{tabular}{|m{2cm}|m{2cm}|m{4cm}|>{\centering\arraybackslash}m{3cm}|}
\hline 
Band  & Symmetry  & Assignment & Virtual NTO\\
\hline 
\textbf{A}  & $A$  & $3s$  & \includegraphics[scale=0.4]{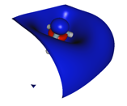}\\
\hline 
\textbf{B}  & $E$  & $3p_{x}$, $3p_{y}$  & \includegraphics[scale=0.4]{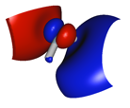}\\
\hline 
\textbf{C}  & $A$  & $3p_{z}$  & \includegraphics[scale=0.4]{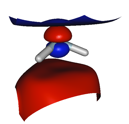}\\
\hline 
\textbf{D}  & $E$  & $3d_{xz}$, $3d_{yz}$  & \includegraphics[scale=0.4]{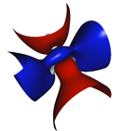}\\
\hline 
\textbf{E}  & $A$  & $4s$  & \includegraphics[scale=0.4]{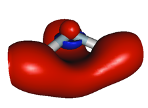}\\
\hline 
\end{tabular}
\end{table}

The XAS for \ce{NH3} as presented in Figure~\ref{fig:XAS-NH3}, shows that there are five absorption bands in the range of 400 to 408 eV. These bands \textbf{A}, \textbf{B}, \textbf{C}, \textbf{D}, and \textbf{E} are due to electronic transition from the N1s orbital to $3s$, $3p$ and $3d$, and $4s$ orbitals (Table~\ref{tab:NH3}). Similar to the case of water, the absorption band due to the $3s$ orbital (band \textbf{A}) for higher electric fields shows a red-shift in the absorption energy as well as suppression of oscillator strength. Likewise, for fields between -0.01 and \SI{0.01}{\au}, and above \SI{0.04}{\au} the bands \textbf{B} and \textbf{C} are well separated, while in the same region as the suppression of band \textbf{A}, these bands again become quasi-degenerate. These features indicate a similar spherical electronic distribution effect as observed in \ce{H2O}.

Absorption bands \textbf{D} and \textbf{E} split with increasing electric field. Band \textbf{D} consists of two degenerate core-excited states into $3d_{xz}$ and $3d_{yz}$ orbitals, whereas band \textbf{E} is formed by excitation to $4s$ orbital. The different trend of the $3s$ (\textbf{A}) and $4s$ (\textbf{E}) energies with field strength may seem anomalous, however the rapid red-shift of band \textbf{A} beyond \SI{0.01}{\au} can instead be explained by mixing with excitation to the $3p_z$ orbital (band \textbf{C}). In fact, these bands completely switch electronic character for fields above $\sim\SI{0.03}{\au}$, which can be seen from the field-dependent virtual NTOs presented in the ESI. The stabilization of the $3p_z$ excitation is then a straightforward effect of the increasing molecular dipole moment.

As with \ce{H2O} and \ce{H2O2}, we find that the excitations to $3d$ orbitals gain intensity for higher electric field strengths. In general, because we find a lengthening of the NH and OH bonds, it is most likely that this results in a lowering of the $\sigma_{XH}^*$ bond energy and enhanced hybridization of the $3d$ orbitals. As these orbitals pick up valence character then naturally they would gain intensity from improved orbital overlap. It is difficult to quantitatively measure such an effect from the NTOs, although some effect may be evident from increasing isosurface volume (see ESI). A more in-depth analysis e.g. based on natural bonding orbitals is outside the scope of this work.

\begin{comment}
We find that at higher electric fields, absorption bands due to excitations to $3s$, $3p$ orbitals undergo intensity depletion, while bands due to excitations to $3d$ and $4s$ orbitals gain intensity. \textbf{A} possible reason for this change intensity could be that at higher electric field strengths the molecular geometry tends toward planar structure, which cause $3d$ and $4s$ orbitals to hybridize with the $\sigma^{*}$orbitals along NH bonds. This can be understood from comparison with the XAS spectra at various electric field strengths, for geometry frozen to field-free optimized structure as presented in Figure S3 of the SI. From this comparison we find that the intensities of the bands \textbf{A}, \textbf{B}, \textbf{D}, and \textbf{E} does not change significantly, which hints that changing intensity is possibly an effect associated with the geometry change. Further, we find that the splitting of \textbf{D} and \textbf{E} bands is caused due to changes in geometry due electric fields. 
\end{comment}

\subsection{Carbon monoxide}

\begin{figure}
\includegraphics[scale=0.35]{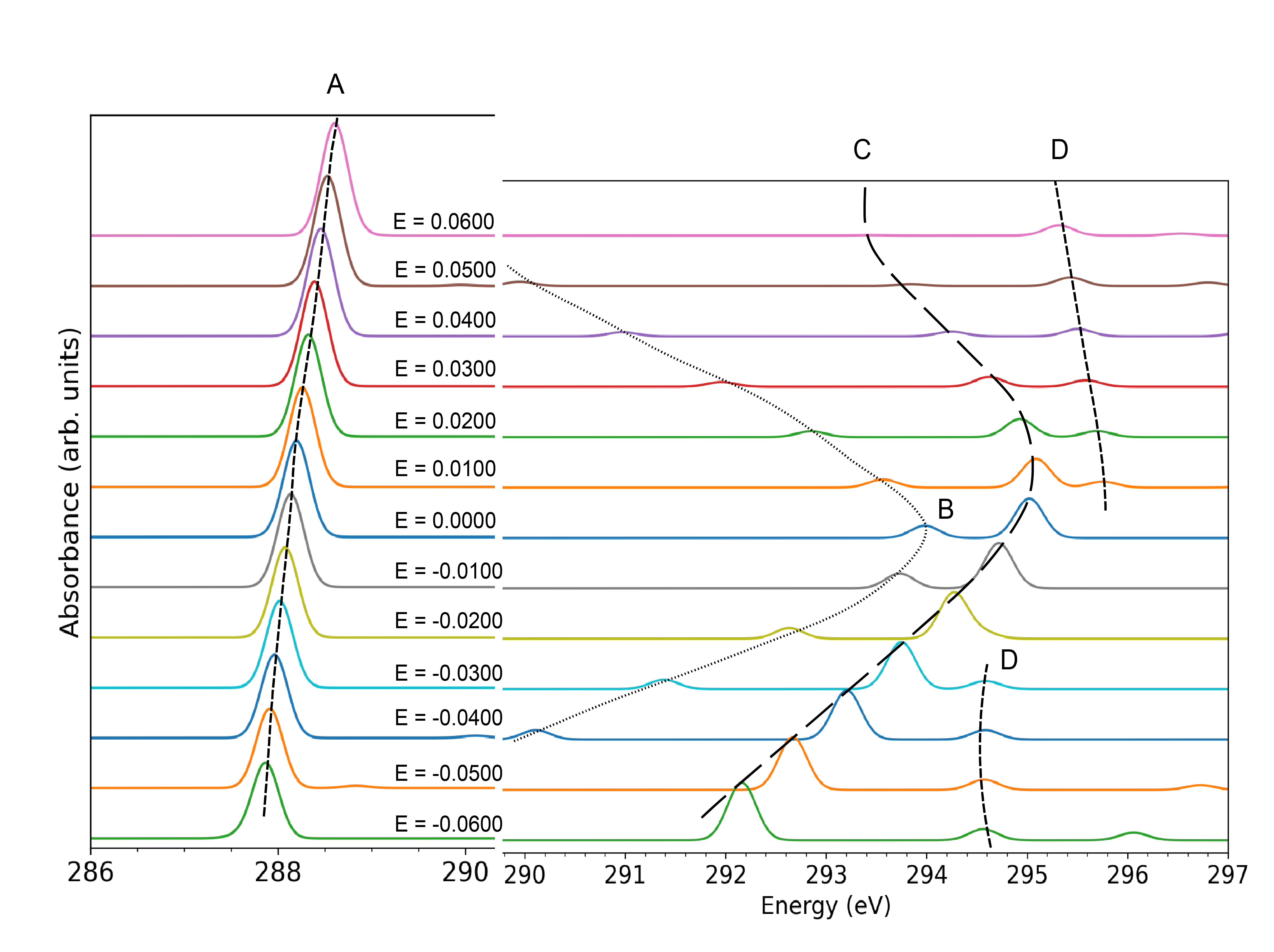}

\caption{\ce{CO}: Evolution of the C1s XAS as a function of EEF strength. Corresponding geometries are optimized in the presence of the electric field. Significant absorption bands are indicated by capital letters. The region from 290--\SI{297}{eV} has been magnified by a factor of 4.2 for clarity.\label{fig:XAS-CO-C}}
\end{figure}

\begin{table}
\caption{\ce{CO}: NTOs of the C1s core-excited states.\label{tab:CO-C}}

\begin{tabular}{|m{2cm}|m{2cm}|m{4cm}|>{\centering\arraybackslash}m{3cm}|}
\hline 
Band  & Symmetry  & Assignment & Virtual NTO\\
\hline 
\textbf{A}  & $\Pi$  & $\pi^{*}$  & \includegraphics[scale=0.4]{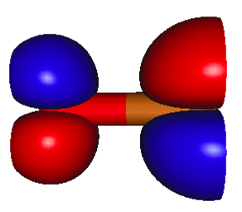}\\
\hline 
\textbf{B}  & $\Sigma$  & $3s$  & \includegraphics[scale=0.4]{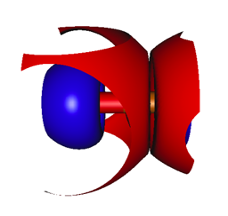}\\
\hline 
\textbf{C}  & $\Pi$  & $3p_{x,y}$  & \includegraphics[scale=0.4]{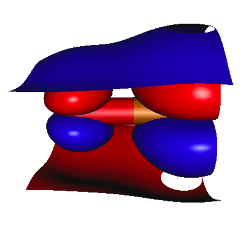}\\
\hline 
\textbf{D}  & $\Pi$  & $3d$/$\pi^*$  & \includegraphics[scale=0.4]{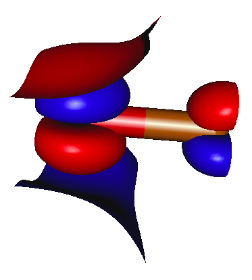}\\
\hline 
\end{tabular}
\end{table}

\ce{CO} shows a much more well-structured C1s XAS response to an EEF (Figure~\ref{fig:XAS-CO-C} and Table~\ref{tab:CO-C}). In this case, the electric field is applied along the bond axis of the \ce{CO} molecule. Increasingly positive field strengths (slightly) elongate the CO bond (Figure~S1) and hence lower the $\pi^*$ orbital energy due to decreased orbital overlap. However, this effect is minor, as evidenced by the fixed-geometry XAS in Figure~S9a. Instead, the main effect is an electronic weakening of the $\pi$ bond which results in a red-shift of band \textbf{A} which is highly linear with field strength. The changes in energy and dipole moment (Figure~\ref{fig:dm-ediff}) are better described as quadratic, as explained by (2) and further differentiation which then involves the anisotropic polarizability. However, the energy \emph{difference} leading to band \textbf{A} relies on the \emph{change} in dipole moment (first-order perturbation) and polarizability (second-order perturbation) upon excitation. The N1s to $\pi^*$ excitation decreases the overall dipole moment, but even more so at increased field strengths (due to increased polarization of the bond), while the polarizability is largely unaffected. This leads to the highly linear shape observed. The large electronic perturbation induced by the EEF is also seen in the atomic partial charges (Figure~S1). Interestingly, the partial charge separation is not linearly dependent on the field, especially above \SI{0.04}{\au}, although this may be an artifact of Mulliken population analysis.

Instead, bands \textbf{B} and \textbf{C} show a nonlinear response. Band \textbf{D}, to the extent that it can be assigned as a single band, shows a less severe field-dependent response, but its low oscillator strength makes an in-depth analysis difficult. These bands originate from Rydberg-type transitions. Band \textbf{B} does not seem to significantly interact with any other states and maintains a consistent electronic character across all field strengths. On the other hand, band \textbf{C} seems to pick up some valence $\pi^*$ character and borrow intensity at negative field strengths. At positive field strengths, this band also moves to lower energy but instead the oscillator strength drops to near zero due to a loss of polarization in the direction of the molecular axis. The parabolic shape of these bands mirrors that of the ground state (Figure~\ref{fig:dm-ediff}). Unlike the $\pi^*$ transition (band \textbf{A}) though, here the dipole moment change is a secondary factor and the increase in polarizability due to the diffuse nature of the Rydberg orbitals drives the red-shift for both positive and negative fields. This same effect may also be at play in the previous systems (\ce{H2O2}, \ce{H2O}, and \ce{NH3}), although $\sigma^*$ mixing, geometric effects, and other perturbations make a clear trend difficult to pinpoint.

\begin{figure}
\includegraphics[scale=0.35]{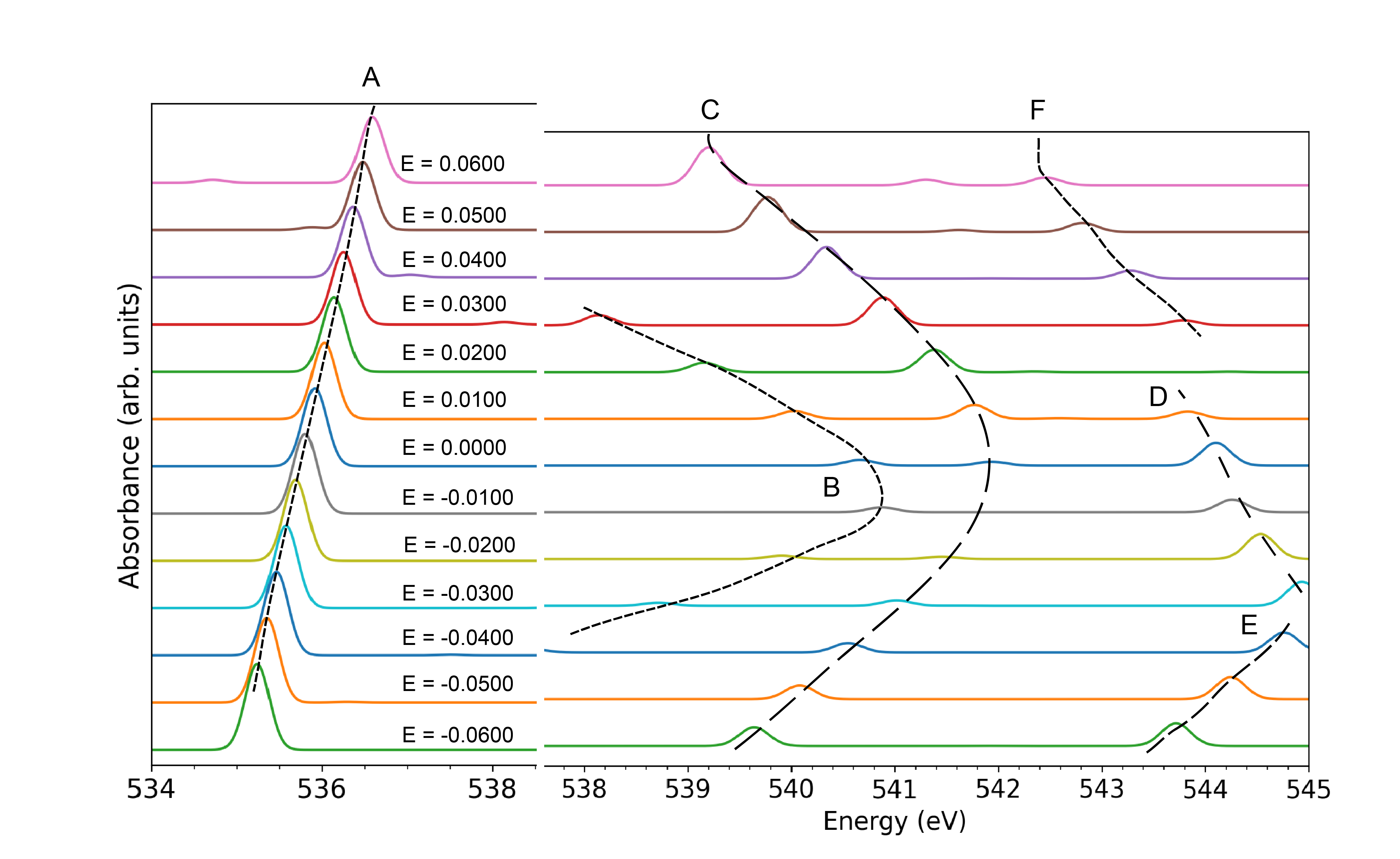}

\caption{\ce{CO}: Evolution of the O1s XAS as a function of EEF strength. Corresponding geometries are optimized in the presence of the electric field. Significant absorption bands are indicated by capital letters. The region from 538--\SI{545}{eV} has been magnified by a factor of 4.5 for clarity.\label{fig:XAS-CO-O}}
\end{figure}

\begin{table}
\caption{\ce{CO}: NTOs of the O1s core-excited states.\label{tab:CO-O}}

\begin{tabular}{|m{2cm}|m{2cm}|m{4cm}|>{\centering\arraybackslash}m{3cm}|}
\hline 
Band  & Symmetry  & Assignment & Virtual NTO\\
\hline 
\textbf{A}  & $\Pi$  & $\pi^*$  & \includegraphics[scale=0.4]{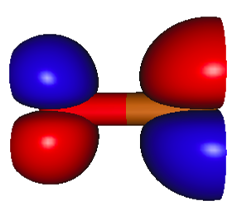}\\
\hline 
\textbf{B}  & $\Sigma$  & $3s$ & \includegraphics[scale=0.4]{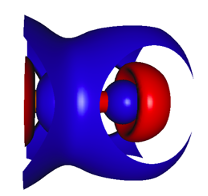}\\
\hline 
\textbf{C}  & $\Pi$  & $3p_{x,y}$  & \includegraphics[scale=0.4]{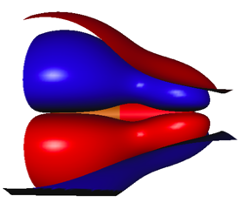}\\
\hline 
\textbf{D}  & $\Pi$  & $3d$/$\pi^*$  & \includegraphics[scale=0.4]{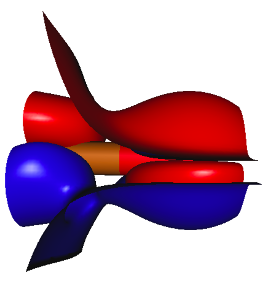}\\
\hline 
\end{tabular}
\end{table}

The O1s XAS of \ce{CO} (Figure~\ref{fig:XAS-CO-O} and Table~\ref{tab:CO-O}) show overall similar trends as for C1s. The most significant exception is band \textbf{C}, which has essentially reverse trends (more significant red-shift and intensity borrowing now at positive field strengths). This points to a possible explanation: red-shifting of band \textbf{C} at either field polarity leads to valence $\pi^*$ mixing, but for intensity enhancement, there must be a transfer of electron density \emph{away} from the core orbital. For O1s, this means that mixing with a highly polarized $\pi^*$ orbital (which is skewed towards the C atom) results in a significant transition dipole moment (TDM), while polarization in the opposite direction does not induce as significant a movement of charge. The precise opposite is the case in C1s XAS, where polarization of the bond towards C (and hence $\pi^*$ towards O) in the negative field direction leads to a large TDM. It is this specificity for highly localized electronic density changes that is hallmark of XAS analysis.

\begin{comment}
Similar to the case of C1s core-excitation XAS, we observe structured response of O1s core-excited states to the applied electric field. The absorption bands for O1s XAS of CO with varying electric field strengths are presented in Figure~\ref{fig:XAS-CO-O}. Simialr to C1s core-excited XAS, we have plotted the O1s XAS for the bands \textbf{B}, \textbf{C}, and \textbf{D} by zooming the spectra (i.e. reducing the spacing along the y-axis and shifting XAS accordingly) in the range of 537.8 eV to 545 eV. Table~\ref{tab:CO-O} presents the symmetry, nature and shape of NTOs for the absorption bands. The O1s core-excited XAS show very similar to C1s core-excited XAS, except we see that the intensities of the bands \textbf{B} and \textbf{C} show increasing intensity at negative electric field strengths. This behavior is opposite to that of C1s core-excited XAS, where bands \textbf{B} and \textbf{C} gain intensity at positive electric fields. Further the band \textbf{D} shows a systematic blue shift in the absorption energy from electric field strengths \SI{-0.01}{\au} to \SI{0.03}{\au} (the band shifts beyond the range of calculated core-excited states for higher electric fields). Further, we observe that the Stark shifts in the O1s spectra have negligible effects of geometry relaxation from comparison with O1s XAS for geometry fixed to field-free optimized geometry at different electric field strengths (Figure S4b of the SI).
\end{comment}

\subsection{Carbon Dioxide}

\begin{figure}
\includegraphics[scale=0.35]{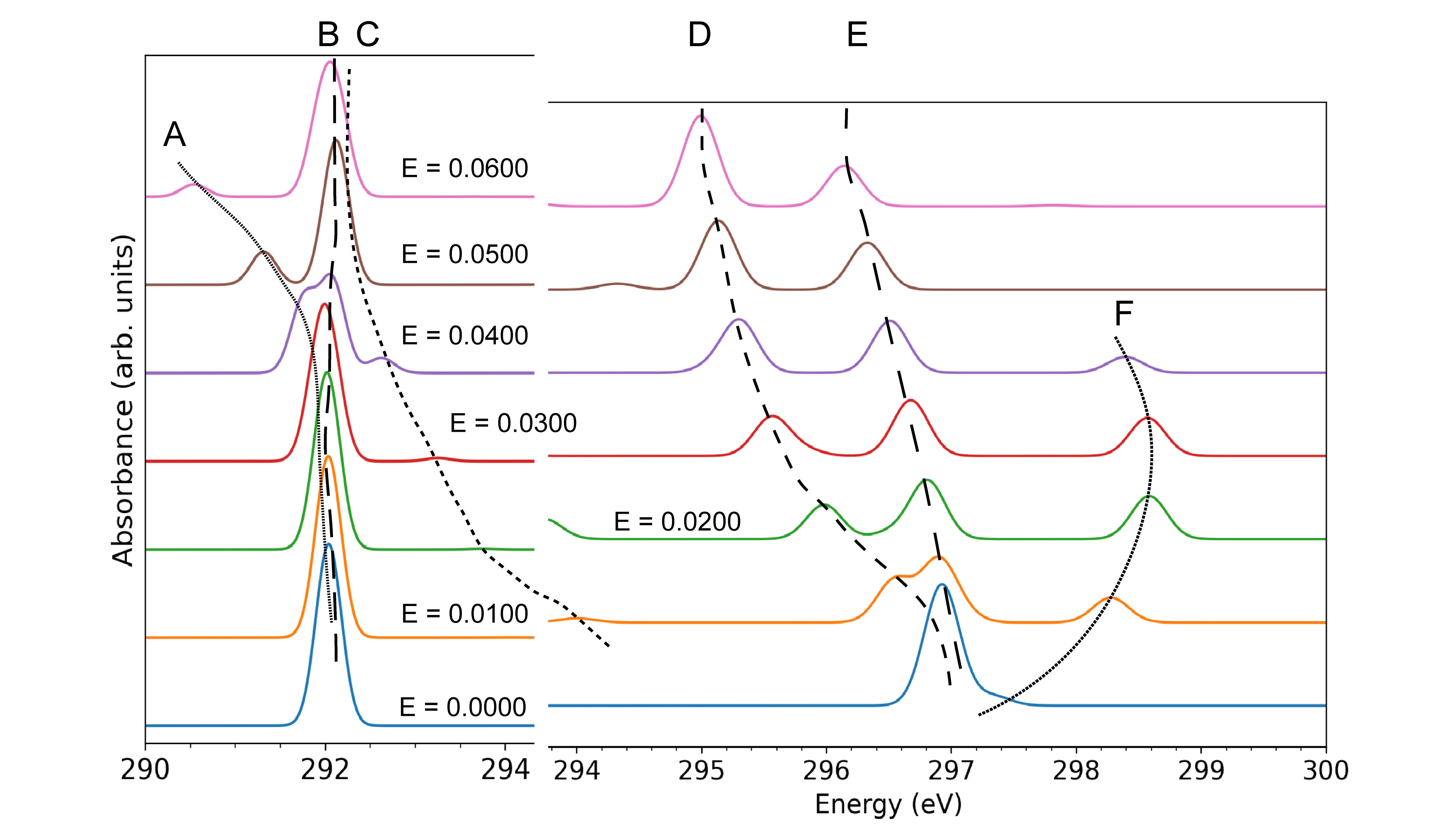}

\caption{\ce{CO2}: Evolution of the C1s XAS as a function of EEF strength. Corresponding geometries are optimized in the presence of the electric field. Significant absorption bands are indicated by capital letters. The region from 294--\SI{300}{eV} has been magnified by a factor of 25 for clarity.\label{fig:XAS-CO2-C}}
\end{figure}

\begin{table}
\caption{\ce{CO2}: NTOs of the C1s core-excited states.\label{tab:CO2-C}}

\begin{tabular}{|m{2cm}|m{2cm}|m{4cm}|>{\centering\arraybackslash}m{3cm}|}
\hline 
Band  & Symmetry  & Assignment & Virtual NTO\\
\hline 
\textbf{A}  & $A_{1}$  & $\pi_{z}^{*}$  & \includegraphics[scale=0.4]{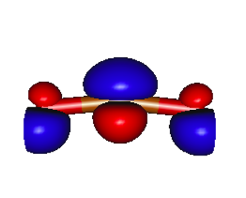}\\
\hline 
\textbf{B}  & $B_{2}$  & $\pi_{y}^{*}$  & \includegraphics[scale=0.4]{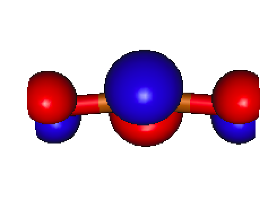}\\
\hline 
\textbf{C}  & $A_{1}$  & $3s$  & \includegraphics[scale=0.6]{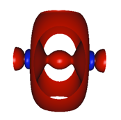}\\
\hline 
\textbf{D}  & $A_{1}$  & $3p_{z}$  & \includegraphics[scale=0.6]{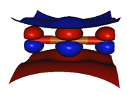}\\
\hline 
\textbf{E}  & $B_{2}$  & $3p_{y}$  & \includegraphics[scale=0.6]{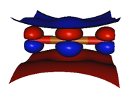}\\
\hline 
\end{tabular}
\end{table}

Figure~\ref{fig:XAS-CO2-C} shows XAS for C1s core-excited states of the carbon monoxide molecule, while Table~\ref{tab:CO2-C} presents the features of the main absorption bands. Due to the molecular symmetry, we have plotted the spectra for the field only in positive direction. Bands \textbf{D}--\textbf{F} have very small oscillator strengths compared to the main peaks. These bands originate from excitation to Rydberg $3p$ and $4s$ orbitals, which are quite weak due to the lack of valence orbital mixing and the high symmetry of the system. Conversely, the weakly-perturbed absorption bands \textbf{A} (low-field)/\textbf{C} (high-field) and \textbf{B} are due to excitation to $\pi_{z}^{*}$ and $\pi_{y}^{*}$ orbitals, respectively (Table~\ref{tab:CO2-C}, note that the electronic character of bands \textbf{A} and \textbf{C} swap around \SI{0.04}{\au}), while the high-field portion of band \textbf{A} and the low-field portion of \textbf{C} are due to excitation to the $3s$ orbital. Note that the symmetry and axes are classified by the perpendicular $C_{2v}$ point group induced by the field perturbation. Band \textbf{B} is essentially unaffected by the EEF as the virtual NTO is exactly perpendicular to the field direction and the excited $3p$ state is not very polarizable in that direction (or rather, exactly as polarizable as in the ground state). The $3p_z$ transition is oriented in the direction of the field and is slightly shifted due to bending of the $\angle$OCO bond angle up to \SI{4}{\degree} (Figures~S5a and c.f. Figure~S10a for the frozen geometry spectrum). The $3s$ Rydberg excitation is red-shifted strongly by the EEF due to enhanced polarizability in the excited state; this band also picks up significant intensity at medium fields strengths due to mixing with the $3p_z$ excitation, which also results in the exchange of electronic character. Band \textbf{D} picks up intensity as it shifts to lower energies at high field strengths, most likely due to mixing with the $\pi^*$ transition of the same symmetry (band \textbf{B}). Note that there is an additional, unassigned peak near \SI{294.3}{eV} for a field strength of \SI{0.05}{\au}. This peak only has non-negligible intensity in this higher-field region; symmetry ($\Sigma_u^+$) and NTO analysis indicate potential assignment as a core to antisymmetric lone pair transition.

\begin{comment}
We noticed that the band \textbf{B} does no show a Stark shift in the absorption energy, while band \textbf{A} splits from the \textbf{B} at electric field beyond \SI{0.04}{\au} This splitting is caused by distortion in the geometry of \ce{CO2} molecule. The C1s XAS at field-free optimized structure for different electric field (as shown in the Figure S5a of the SI) shows no splitting of \textbf{A} and \textbf{B} absorption bands, further both of these bands show no shift in the absorption energy as the electric field strength is changed. The absorption band \textbf{C}, which is caused by excitation to the $3s$ orbital, shows a systematic red-shift in the absorption energy. This red-shift in the absorption energy is similar to that in the core-excitation to $3s$ orbitals in the other molecules studies here. The absorption bands \textbf{D} and \textbf{E} undergo splitting in the presence of electric field, this splitting in absorption energy further increases as the field intensity is increased due to hybridizing of $3p_{x}$ orbital. 
\end{comment}

\begin{figure}
\includegraphics[scale=0.35]{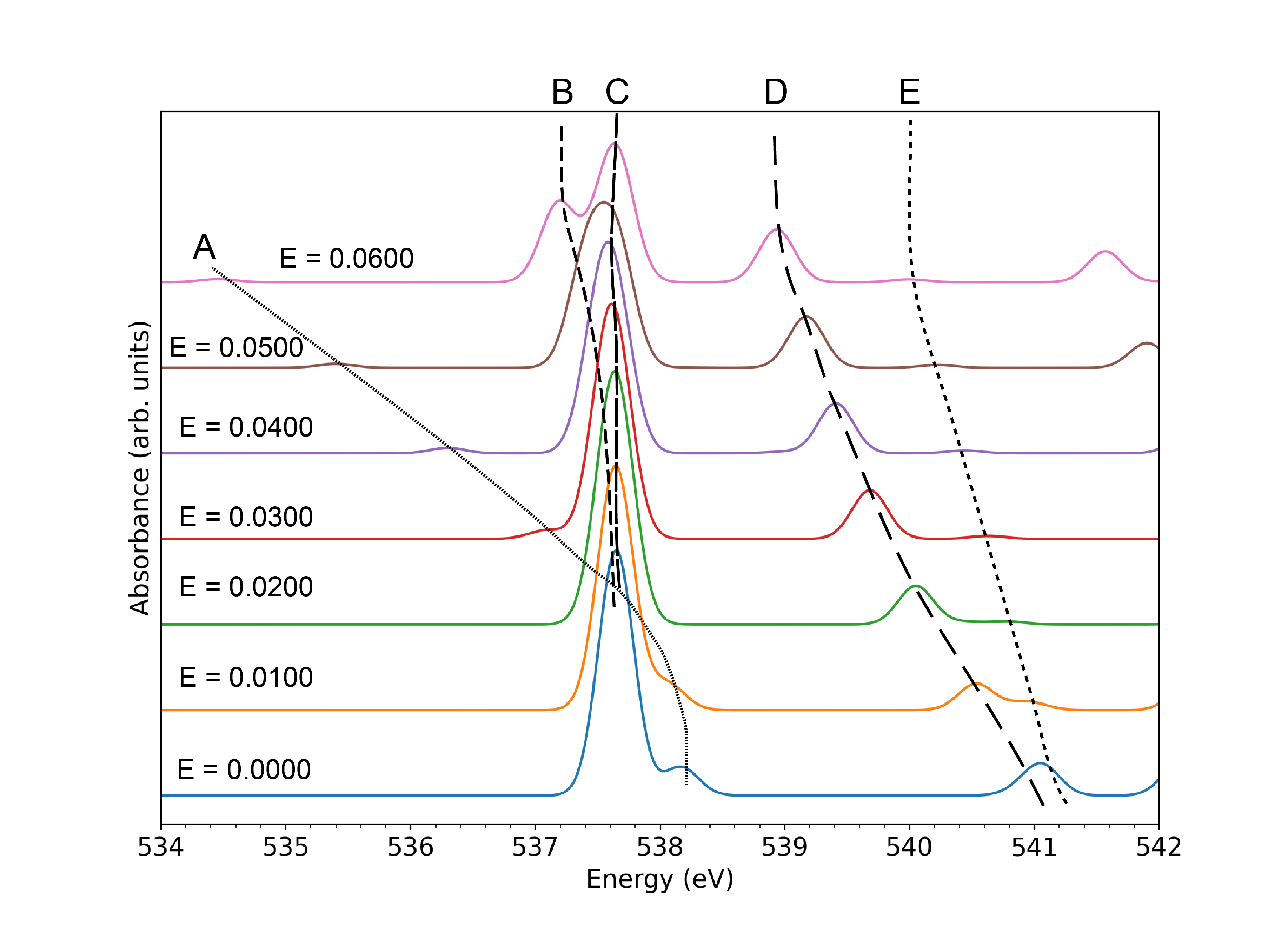}

\caption{\ce{CO2}: Evolution of the O1s XAS as a function of EEF strength. Corresponding geometries are optimized in the presence of the electric field. Significant absorption bands are indicated by capital letters.\label{fig:XAS-CO2-O}}
\end{figure}

\begin{table}
\caption{\ce{CO2}: NTOs of the O1s core-excited states.\label{tab:CO2-O}}

\begin{tabular}{|m{2cm}|m{2cm}|m{4cm}|>{\centering\arraybackslash}m{3cm}|}
\hline 
Band  & Symmetry  & Assignment & Virtual NTO\\
\hline 
\textbf{A}  & $A_{1}$  & $3s$  & \includegraphics[scale=0.6]{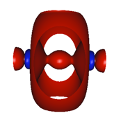}\\
\hline 
\textbf{B}  & $A_{1}$  & $\pi_{x}^{*}$  & \includegraphics[scale=0.6]{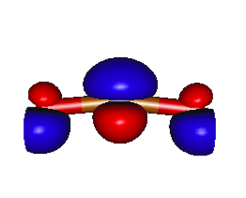}\\
\hline 
\textbf{C}  & $B_{2}$  & $\pi_{y}^{*}$  & \includegraphics[scale=0.5]{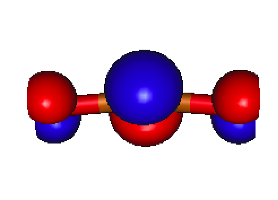}\\
\hline 
\textbf{D}  & $A_{1}$ & $3p_{z}$  & \includegraphics[scale=0.6]{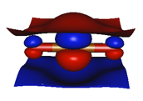}\\
\hline 
\textbf{E}  & $A_{1}$ & $4s$  & \includegraphics[scale=0.5]{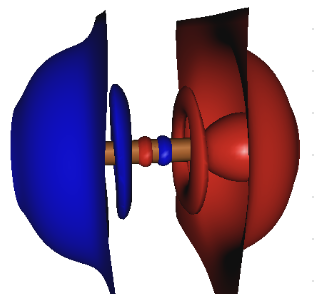}\\
\hline 
\end{tabular}
\end{table}

The low-energy region of the O1s XAS for \ce{CO2} (Figure~\ref{fig:XAS-CO2-O}, Table~\ref{tab:CO2-O}) shows similar trends as for C1s; although one primary difference is that, as for \ce{H2O2}, there are two O1s molecular orbitals of different symmetry. For example, this provides for a symmetry-allowed transition to the $3s$ orbital at zero field (band \textbf{A}), which also occurs at relatively lower energy as for C1s (measured from the primary $\pi^*$ absorption peaks, bands \textbf{B} and \textbf{C}). As before, this band red-shifts at higher fields and loses intensity above \SI{0.04}{\au}. Energetically, the O1s XAS is quite complex at higher energies, with a number of states red-shifting dramatically and encroaching on the $\pi^*$ excitations. This includes excitations to the $3p$ orbitals but also $4s$ and some $3d$ orbitals. However, none of these excitations, save one (band \textbf{D}, excitation to $3p_z$ at zero field) accrue any appreciable intensity. In spite of this trend, band \textbf{D} does in fact gain significant intensity (becoming almost as intense as band \textbf{B}) at higher fields. However, this band switches electronic character with band \textbf{E} (excitation to the $4s$ orbital) at field strengths between 0 and \SI{0.01}{\au}. We attribute this primarily to mixing with band \textbf{B}, but also to the polarization of the $3p_z$ orbital towards the C atom, which is also incorporated into the $4s$ character due to mixing. Both the $3s$ and $4s$ Rydberg excitations are significantly stabilized in this spectrum compared to the C1s spectrum. This can be explained by the better spatial overlap of the terminal O1s core orbitals with the diffuse Rydberg orbitals (note that they do \emph{not} overlap well with the $\pi^*$ or $3p$ orbitals due to $p$-type nodal structure on the O atoms).

\begin{comment}
The XAS profile for O1s core-excited states of CO$_{2}$molecule is plotted in the range of 534 eV to 542 eV, as presented in the Figure \ref{fig:XAS-CO2-O}. The absorption band \textbf{A}, result of excitation to $3s$ orbital (Table~\ref{tab:CO2-O}), undergoes systematic red-shift in the absorption energy as the electric field intensity is increased. This observation is similar to C1s XAS as discussed earlier. Also, similar to the case of C1s XAS, the absorption bands \textbf{B} and \textbf{C} (excitation of C1s to $\pi_{x}^{*}$ and $\pi_{y}^{*}$ orbitals) split at higher electric field strengths. It is interesting to note that, unlike C1s XAS, for the O1s XAS at geometry fixed to field-free optimized structure (Figure S5b of the SI) as well the bands \textbf{B} and \textbf{C} split at higher electric field strengths. The absorption band \textbf{D}, which is a result of excitation to $3p_{y}$ orbital, shows a systematic red-shift in the absorption energy and increasing intensity as the electric field strength is increased. The increase in the oscillator strengths at higher electric fields is caused by hybridization of $3p_{y}$ orbital with a $\pi^{*}$ type orbital. 
\end{comment}

\section{Discussion}

The core orbitals are highly localized and hence they are sensitive to the local electronic environment. This environment is sensitive to applied electric fields via changes in local electronic structure (polarization, charge separation, orbital energy perturbations, etc.), as well as geometry relaxation. Thus, x-ray spectroscopy should be an excellent tool to study Stark effects in molecules at a detailed, sub-molecular level. Our XAS calculations also suggest a vital role of the virtual valence/Rydberg orbitals, in addition to the core orbitals. 

As a result of applied electric field, the excitation energy of the core-excited state as well as the corresponding oscillator strength are affected. The response of the energy and oscillator strength seem to provide very different but highly complementary information, with the former driven by the electronic structure of the virtual orbitals, and the latter driven by highly local interaction of the core orbital with the delocalized virtual orbital space.

\begin{comment}
We observed that the core-excited states involving excitations to $3s$ orbitals show higher sensitivity towards absorption energy and the oscillator strength with varying electric field strength. This could be a result of additional (or reduction in) partial charge on the atom, whose core-excitation is probed, at higher electric field strengths. A similar effects on the oscillator strengths of $3p$ orbitals is observed. In general it is observed that the core-excited states involving $3p$ orbitals show a splitting into their absorption bands at higher field strengths, irrespective of geometrical relaxation. A similar splitting is observed for bands involving excitations to $3d$ orbitals (in some cases merged with transitions to $4s$ orbitals). Further, we observed that the intensities of the transitions to the valence orbitals are less sensitive to the electric field strength and splitting in such bands is caused if asymmetric geometrical distortions are made due to application of electric field. 
\end{comment}

Three types of Stark shifts in the absorption energies are observed: linear, non-linear and none. The excitations to valence orbitals typically show a linear Stark shift, typified by the absorption band \textbf{A} of C1s XAS for CO (Figure~\ref{fig:XAS-CO-C}). This shift arises due to perturbation of bonds which lie parallel to the applied field direction, with contributions from increased (or decreased) polarization as well as modification of the bond length which decreases the anti-bonding orbital energy. Based on comparison of relaxed and frozen geometry calculations (see ESI), the former effect seems to be dominant especially in multiply-bonded species.

A non-linear Stark shift is observed most strongly in transitions to Rydberg orbitals, mostly $3s$ and $3p$ states in our calculations. For example the absorption bands \textbf{B} and \textbf{C} in the CO spectra (Figures~\ref{fig:XAS-CO-C} and \ref{fig:XAS-CO-O}) and band \textbf{A} in the \ce{CO2} O1s spectrum (Figure~\ref{fig:XAS-CO2-O}). This effect is due to the increased polarizability in the excited state (due to its more diffuse character), which creates a strong, non-linear (quadratic) energetic sensitivity of the energy difference to the EEF strength via differentiation of (2).

The XAS of \ce{H2O} and \ce{NH3} exhibit a mixture of these effects, as the spectra include essentially only Rydberg-type transitions, but the molecular geometry also features polar OH or NH bonds oriented partly along the field direction. This leads to excitation profiles with a shifted parabolic shape, see for example band \textbf{A} of O1s XAS for \ce{H2O2} (Figure~\ref{fig:XAS-H2O2}), band \textbf{A} in O1s XAS for water (Figure~\ref{fig:XAS-H2O}), and band \textbf{A} of N1s XAS for \ce{NH3} (Figure~\ref{fig:XAS-NH3}). In these cases, the hydrogen atom (of partial positive charge at zero field) is oriented "against" the positive field direction, or in other words, the bond dipoles are aligned with the overall molecular dipole moment. This leads to a vertical excitation energy which peaks at positive field strength (which is the easiest to apply experimentally given freely-orienting molecules). Conversely, the opposite trend should be obtainable in more complex molecules where the local bond dipole is in opposition to the overall molecular dipole moment and hence EEF direction.

Finally if the electronic transition occurs to an orbital that is oriented perpendicular to the applied electric field then little or no shift is observed for the transition energy with the applied field strength. For example, in the case of \ce{CO2} the absorption band \textbf{B} in C1s XAS (Figure~\ref{fig:XAS-CO2-C} and the absorption band \textbf{C} in O1s XAS (Figure~\ref{fig:XAS-CO2-O}). Together, these differing energetic responses provide information on electronic orientation, valence vs. Rydberg character, as well as bond polarity, in addition to more subtle effects not fully analyzed here.

In addition to this energetic information, the oscillator strengths of individual bands provide more specific information related to the core orbitals and their location within the molecular structure. A prime example is seen in band \textbf{C} of \ce{CO} XAS: in the C1s spectrum (Figure~\ref{fig:XAS-CO-C}), this band is strongly attenuated at positive fields strengths, but is moderately intense at zero field, and even enhanced at negative field strengths. A nearly opposite trend is observed in the O1s spectrum (Figure~\ref{fig:XAS-CO-O}), where little intensity is evident at zero field, but positive field strengths lead to intensities to those observed for C1s. This effect is due to the relationship between the movement of charge density from ground to excited state, encoded neatly by the virtual NTO(s), and the position of the core orbital. In this case (as with band \textbf{D} of \ce{CO2} O1s, Figure~\ref{fig:XAS-CO2-O}), the intensity seems to be driven at least in part by a movement of charge from the core region towards adjacent atoms. In larger systems, the effect or core orbital locality should be even more pronounced as the "area" covered by each core orbital shrinks in relation to the total molecular size, and a wider variety of electronic excitations are encountered. We expect this feature to be especially relevant in the disambiguation of multiple $\pi^*$-type excitations which are often energetically overlapped, but spatially restricted to particular double or triple bonds.

The Stark effect on the singlet electronic transitions in small molecules is studied by Gurav et. al.\cite{Gurav2018} This study points out that the UV absorption bands undergo shifting, merging, and splitting as the strength of the EEF is changed. While this result is similar to our observation of shifts in core-excitation energies in x-ray absorption spectra, the shifts in the core-excited states seem to be more systematic than in the valence excited states. The variation in the transition dipole moment due to EEF for valence electronic excitations depends on the relative orientations of the occupied and virtual orbitals. For valence-excited states, this variation in the relative orientation of the orbitals is non-trivial, while for the K-edge core-excited states it mostly depends on the changes in the virtual orbital orientations. The electronic Stark effect in the UV region is useful for studying the phenomena that occur on the excited states potential energy surfaces, such as photocatalytic water splitting.\cite{Bachler2016} On the other hand, the soft x-ray Stark effect is promising as an attractive tool for understanding the mechanistic picture of the ground state electronic structure and the its perturbation by the local chemical environment.

Currently, vibrational Stark spectroscopy is widely used to for quantifying various weak interactions in terms of electric fields.\cite{Fried2015} It is possible that x-ray Stark spectroscopy can provide complementary results to vibrational Stark spectroscopy. Further, element-selective features of x-ray spectroscopies could be exploited for studying the local electric fields exerted due to weak interactions in biological systems. Additionally, as there growing interests in using external electric fields as catalysts, in that regard the x-ray Stark effect can be proven useful to understand chemical selectivity as a result of electric fields. 

\section{Conclusions}

We have investigated the effects of electric fields on the x-ray absorption spectra of a set of small prototypical molecules. Individual, field-dependent absorption bands were identified based on the electronic transitions to various excited orbitals based on NTO analysis. Further, to understand the effects of geometrical relaxation and response from electron density to the applied electric fields, we calculated the XAS for geometries frozen to the corresponding field-free optimized structures.

Our analysis of the computed spectra leads to a number of conclusions about the information which might be obtainable from such spectra:
\begin{enumerate}
    \item Excitations to valence orbitals (especially $\pi^*$ orbitals) tend to lead to linear shifts in absorption energy with applied field. Fields applied in the same direction as the bond dipole lead to a raising of the absorption energy.
    \item Excitations to Rydberg orbitals tend to lead to quadratic changes in energy, with absorption energy decreasing for both positive and negative field strengths. The magnitude of this effect depends on the change in polarizability from the ground to excited state.
    \item The magnitude of energetic shifts depends strongly on the orientation of the virtual orbital (e.g. as exemplified by the virtual NTO) with respect to the applied field.
    \item Geometric relaxation can have a significant effect on the absorption spectra. The effect of changes in various geometric parameters is highly non-uniform and seems to follow the general trend: dihedral angles (especially single-bond rotations with polar substituents) $\gg$ bond angles $>$ bond lengths.
    \item Changes in absorption intensity (oscillator strengths) are driven by local interaction with the core orbital, as well as electronic mixing as a function of energy. Intense transitions require a significant transition dipole moment and orbital overlap with the core orbital. This effect provides element- and regio-specificity.
\end{enumerate}

We hope that this initial study provokes a wider interest in x-ray Stark spectroscopy as a tool to study a wide variety external electronic perturbations using the simple and universal language of electric fields. We also look forward to experimental finite-field XAS data which can corroborate our theoretical findings.

\begin{acknowledgement}
All calculations were performed on the ManeFrame II computing system at SMU. This work was supported in part by the Robert A. Welch foundation under grant no. N-2072-20210327. CW was supported as a Hamilton Undergraduate Research Scholar.
\end{acknowledgement}

\begin{suppinfo}
Electronic supplementary information files are available free of charge at the publisher's website. The following files are included:
\begin{itemize}
\item Supporting\_Information.docx: Computational details, analysis of ground state energy curves, relaxed geometry and atomic charges as a function of field strength, and XAS plots with varying electric field for the frozen (field-free) geometry. 
\item Supporting\_Information\_Raw\_Data.xlsx: optimized geometries, dipole moments, electronic energies and electronic energy differences for various electric fields, core-excited vertical excitation energies and corresponding oscillator strengths, leading NTOs. 
\end{itemize}
\end{suppinfo}

\bibliography{achemso-demo}

\end{document}